\documentclass{article}

\usepackage[preprint]{neurips_2024}

\usepackage{graphicx}
\usepackage{multirow}
\usepackage{float}
\usepackage[figuresright]{rotating}

\usepackage[utf8]{inputenc} 
\usepackage[T1]{fontenc}    
\usepackage{hyperref}       
\usepackage{url}            
\usepackage{booktabs}       
\usepackage{amsfonts}       
\usepackage{nicefrac}       
\usepackage{microtype}      
\usepackage{xcolor}         
\usepackage{threeparttable}
\usepackage{caption}
\usepackage{tabularx}
\newcolumntype{L}{>{\RaggedRight\hangafter=1\hangindent=0em}X}

\usepackage{amssymb}
\makeatletter

\newcommand{\Rmnum}[1]{\expandafter\@slowromancap\romannumeral #1@}
\makeatother

\title{Student-AI Interaction in an LLM-Empowered Learning Environment: A Cluster Analysis of Engagement Profiles}

\author{%
  Zhanxin Hao \\
  School of Education\\
  Tsinghua University\\
  Beijing, China \\
  \texttt{zhanxin\_hao@mail.tsinghua.edu.cn} \\
  \And
  Jianxiao Jiang \\
  School of Education\\
  Tsinghua University\\
  Beijing, China \\
  \texttt{jjx23@mails.tsinghua.edu.cn} \\
  \AND
  Jifan Yu \\
  School of Education\\
  Tsinghua University\\
  Beijing, China \\
  \texttt{yujifan@tsinghua.edu.cn} \\
  \And
  Zhiyuan Liu \\
  Department of Computer Science and Technology \\
  Tsinghua University\\
  Beijing, China \\
  \texttt{liuzy@tsinghua.edu.cn} \\
  \And
  Yu Zhang\thanks{Corresponding author. Address: 
417 Wennan Building, Tsinghua University, Beijing, China, 100084.
Email: zhangyu2011@tsinghua.edu.cn} \\
  School of Education\\
  Tsinghua University\\
  Beijing, China \\
  \texttt{zhangyu2011@tsinghua.edu.cn} \\
}

\begin{document}

\maketitle

\begin{abstract}
  Integrating Large Language Models (LLMs) into educational practice enables personalized learning by accommodating diverse learner behaviors. This study explored diverse learner profiles within a multi-agent, LLM-empowered learning environment. Data was collected from 312 undergraduate students at a university in China as they participated in a six-module course. Based on hierarchical cluster analyses of system profiles and student-AI interactive dialogues, we found that students exhibit varied behavioral, cognitive, and emotional engagement tendencies. This analysis allowed us to identify two types of dropouts (early dropouts and stagnating interactors) and three completer profiles (active questioners, responsive navigators, and lurkers). The results showed that high levels of interaction do not always equate to productive learning and vice versa. Prior knowledge significantly influenced interaction patterns and short-term learning benefits. Further analysis of the human-AI dialogues revealed that some students actively engaged in knowledge construction, while others displayed a high frequency of regulatory behaviors. Notably, both groups of students achieved comparable learning gains, demonstrating the effectiveness of the multi-agent learning environment in supporting personalized learning. These results underscore the complex and multifaceted nature of engagement in human-AI collaborative learning and provide practical implications for the design of adaptive educational systems.
\end{abstract}

\section{Introduction}

The rapid development of large language models (LLMs) has brought transformative changes to education, especially in the design and implementation of learning environments. Traditionally, online learning environments such as Massive Open Online Courses (MOOCs) have provided learners with broad access to vast educational resources and the flexibility of self-paced learning, breaking down the barriers of time and space. However, they often lacked the synchronous, dynamic interactions between student and instructors, as well as among peers. The personalized guidance was also limited as teacher predominantly relied on one-way instructional methods. In recent years, the rise of LLMs has shown the potential to overcome these limitations inherent in traditional online environments. A common practice is using generative AI chatbots to answer student questions. For instance, Khan Academy has integrated LLMs into its platform, allowing students to engage with AI agents for interactive Question \& Answer sessions and to explore newly learned concepts more deeply. Going a step further, AI applications are now supporting and re-constructing the entire educational ecosystem. For example, Google’s Gemini for Education can help teachers create personalized lecture scripts at different difficulty levels and even generate instructional videos from outlines and transcripts for students to watch. These LLM-empowered environments are reshaping the learning paradigm, paving the way for more engaging, personalized, and adaptive learning experience. 

In the new online learning environments reshaped by LLMs, learners are exhibiting novel behaviors and characteristics. Given the inherent capabilities of LLMs in facilitating educational dialogue, a deeper understanding of student interactions with AI agents, particularly through dialogue, can yield new insights into the learning process. Recent studies have begun to analyze the dynamics and varying effects of student-AI interactions, leading to a substantial body of knowledge (e.g., \citealt{guner2025ai, kregear2025analysis, zhu2024human}). However, current research predominantly investigates scenarios where AI serves as a single question-and-answer tool. To our knowledge, there is a distinct lack of research focusing on student engagement within authentic learning environments driven by multiple, specialized LLM agents. This study addresses this gap by investigating student engagement within a Massive AI-empowered Course (MAIC) system, where multiple AI agents provide course instruction and real-time feedback. We aim to explore student learning engagement across a university-level course comprising six modules. This study aims to address the following research questions:

\textbf{RQ1}. How do students engage with the LLM-empowered course?

\textbf{RQ2}. What distinct interaction patterns characterize student dialogues with multiple AI pedagogical agents?

\textbf{RQ3}. How do different engagement patterns relate to learning outcomes and affective changes?

\section{Literature review}

\subsection{Identifying patterns of student engagement in online learning}

Online learning provides flexible study time and abundant learning resources, greatly extending the scale of education. A key consensus from previous research is that the effectiveness of learning largely depends on students’ engagement \citep{means2009evaluation, martin2018engagement}. Thus, understanding student different engagement types and their corresponding behaviors can provide valuable insights into how students process learning content, interact with peers, and utilize various learning resources. Previous researches have explored engagement patterns in online learning from different perspectives and using various methods \citep{kew2022learning}. Some studies utilized qualitative methods, such as interviews and content analysis to categorize students’ engagement types (e.g., \citealt{milligan2013patterns, isda2019study}). In addition, data-driven methods like clustering and other advanced machine-learning algorithms have been employed to identify the status and features of student engagement in online settings \citep{li2022unfolding, liu2022automated}, using various data sources such as behaviors across various activities, time usage, discussion texts, course completion status and so on.

Existing research has identified that student engagement patterns in online learning are highly complex and context-dependent. Across the literature, researchers have broadly categorized these patterns into two primary streams. The first stream addresses the overall level of engagement among student groups, such as “more engaged” and “less engaged”, or “completers” and “non-completers”. For instance, \citet{nkomo2021student} clustered the behaviors of 54 students in a blended learning environment and identified three types of students: high engagement, low engagement, and moderate engagement. Students within these groups demonstrated varying levels of login, posting, and viewing behaviors that corresponded to high, moderate, and low engagement, respectively. Similarly, \citet{alzahrani2025identifying} found similar patterns among students in a 13-week e-learning course, where activities such as attending lectures, submitting assignments, and participating in discussions were distributed proportionally from high to low engagement.

The stream highlights behavioral preferences among different student groups. For example, \citet{liang2008exploration} conducted a study on 60 students using an online learning system, collecting data on their access times to handbooks, exercises, forums, and other course materials. Participants were classified into three categories: active, enthusiastic, and lower participants. Notably, active participants displayed the least preference for handbooks compared to the other groups, and their overall engagement level was intermediate. Similarly, \citet{wise2013broadening} investigated the viewing, posting, revisiting, and editing behaviors of 95 participants in online discussion sessions and categorized them into three distinct types: superficial listeners and intermittent talkers; concentrated listeners and integrated talkers; and broad listeners and reflective talkers. Despite their differing preferences for listening and talking during online discussions, students in three groups exhibited comparable overall performance.

However, the two streams of studies mentioned above are limited to a holistic and explicit perspective, making it difficult to capture students implicit cognitive and motivational processes. In recent years, research has increasingly sought to uncover the internal dynamics of student engagement by integrating behavioral, cognitive, and emotional dimensions. For example, \citet{deng2020linking} analyzed questionnaire data from 1,452 MOOC learners, considering behavioral, cognitive, emotional, and social participation factors, and identified three types of students: “Individually Engaged,” “Least Engaged,” and “Wholly Engaged”, while learning motivation of the three groups of students were significantly different. Similarly, \citet{liu2025exploring} investigated students’ cognitive and emotional engagement in MOOC discussions using deep learning and topic modeling, revealing that increasing topic complexity triggered dynamic shifts in students’ cognitive and emotional responses. Learners engaged in higher-order thinking tended to experience less positive emotion and more confusion or negativity, but those who sustained both positive emotions and complex cognition achieved superior academic outcomes. These studies highlight the importance of adopting multi-faceted perspectives to gain a comprehensive understanding of online student engagement. Building upon this line of research, the present study analyzes students’ engagement profiles based on various data sources and multifaceted dimensions including both behavioral, cognitive and emotional status. This approach not only reveals more nuanced behavioral trends but also offers novel insights into students’ underlying psychological processes and explanatory mechanisms.

\subsection{Student engagement in interactions with AI}

AI technologies, particularly generative AI, are driving significant changes in online learning environments \citep{bozkurt2023challenging}. These tools enable more immediate, efficient, and personalized student interactions across a variety of contexts, such as programming education \citep{boguslawski2025programming}, language learning \citep{creely2024exploring}, and medical education \citep{preiksaitis2023opportunities}. Numerous studies have reported positive outcomes associated with AI integration in education (e.g., \citealt{sun2024does}). However, it is important to note the heterogeneity in these effects—AI does not benefit all learners to the same extent \citep{ma2025meta}. For instance, \citet{dai2025students} found that only students at the highest and lowest achievement levels had significant gains from the AI–assisted physics tutoring, while those in the middle level did not benefit equivalently. These findings highlight the need to further investigate varied student engagement profiles and identify the characteristics that shape them in AI-assisted learning environments, which can provide valuable insights for designing more equitable and personalized AI-driven educational systems, ultimately promoting learning outcomes and advancing research in educational technology.

Recent studies have begun to explore student engagement in AI-empowered learning environments \citep{khosravi2025generative}, reporting results in diverse educational contexts, such as writing tasks (e.g. \citealt{zare2025generative}), project-based learning (e.g. \citealt{perifanou2025collaborative}), STEM (e.g. \citealt{naman2025analysis}), etc. However, most existing research adopts descriptive approaches. For instance, \citet{kregear2025analysis} categorized students’ interactions with LLMs in a physics laboratory by analyzing interaction transcripts and survey responses. They identified "Answer Verification" and "Conceptual Inquiries" as the two most frequent types of student-LLM messages. Students tended to use the LLM to verify their answers or inquire concept interpretation. Similarly, \citet{zhu2024human} examined modes of collaboration between students and generative AI through surveys and reflection questions, revealing three leading patterns: even contribution, human-led, and AI-led interactions. Different leading patterns may indicate students’ distinct engaging approaches, including collaboration, dominance or passive roles in interactions with AI. 

A smaller but growing body of work employs data-driven methods to uncover engagement patterns. For example, \citet{nguyen2024human} utilized Hidden Markov Models (HMM) and sequence clustering to analyze doctoral students’ interactions with ChatGPT during academic writing tasks. Their findings revealed two distinct engagement types: “Structured Adaptive” and “Unstructured Streamline.” Students exhibiting Structured Adaptive patterns demonstrated more reflective behaviors, such as searching for relevant literature and critically editing the AI-generated content, whereas those with Unstructured Streamline patterns tended to simply copy and paste content from generative AI tools. Likewise, \citet{liu2024collaboration} applied sequence analysis and Epistemic Network Analysis (ENA) to examine students’ behaviors while collaborating with generative AI to refine instructional design schemes. The study found that high-performing students engaged in more critical uses of AI-generated content, such as posing exploratory questions and actively modifying the output.
Despite the rapid development of AI-empowered learning environments, our understanding of the underlying interaction patterns and student engagement modes remains limited. Existing research provides only a partial and often descriptive perspective on how students interact with intelligent systems. Therefore, this study aims to analyze student-system interactions, with a particular focus on the semantic content of dialogues between students and AI agents. This research will uncover patterns of student engagement and examines their relationship with learning performance. This novel perspective contributes to a deeper understanding of student-AI interaction and highlights the potential impact of diverse engagement patterns on educational outcomes.

\section{Methods}

\subsection{Settings: Massive AI-empowered Course System (MAIC)}

\begin{figure}[H]
  \centering
  \includegraphics[width=\textwidth]{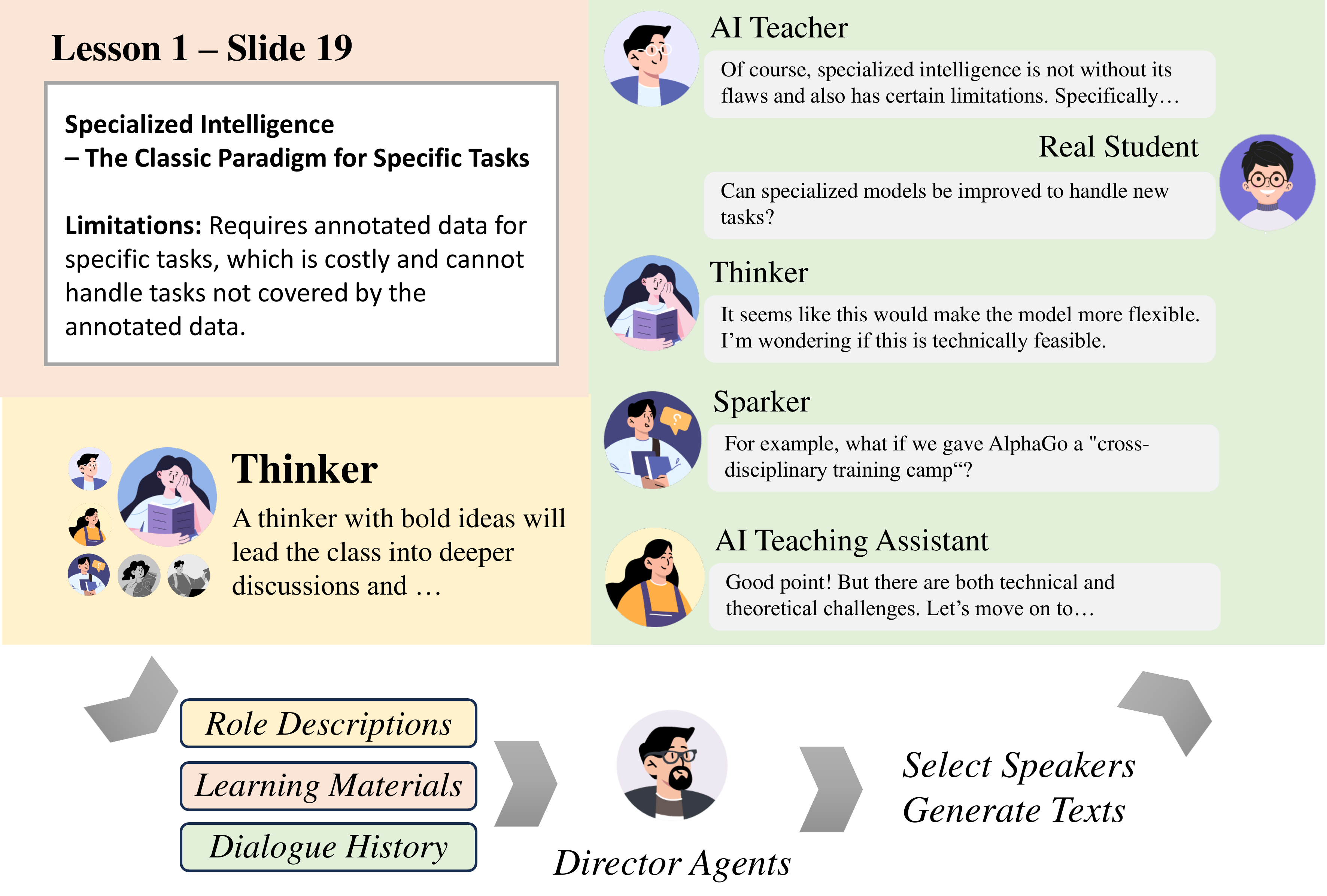}
  \caption{MAIC settings.}
  \label{maic_settings}
\end{figure}

The Massive AI-empowered Course System (MAIC) system is an online learning platform containing a series of LLM-driven agents to support both teaching and learning \citep{yu2024mooc}. Human teachers and teaching assistants collaborate with LLM agents to design courses, including creating teaching slides and generating lecture scripts. Students can view the slides (shown in the upper-left interface of Figure~\ref{maic_settings}) and interact with AI agents through a messaging interface (upper-right in Figure~\ref{maic_settings}). The platform provides several specialized AI agents, including AI Teachers to deliver lectures and answer questions, AI Teaching Assistants to maintain classroom order, and several AI students with different characteristics. As shown in the lower section of Figure~\ref{maic_settings}, the system also includes director agents, who analyze classroom processes and learning conditions based on current learning materials, student-AI dialogue records, and the characteristics of AI agents. Based on the analysis, director agents make decisions about the pace of the class and determine which AI agent should interact with students at specific time.

The MAIC system fosters a collaborative and supportive learning environment, particularly when students facing challenges on specific knowledge areas or tasks. When a student poses a question, multiple agents respond from different perspectives and generate instructional follow-up questions based on students’ inquiries. Additionally, the AI agents provide emotional support by affirming and encouraging students’ ideas. Exemplar student-AI interactive dialogue is illustrated in Table~\ref{eg-dialogue}.

\begin{table}[]
\caption{Student-AI interaction dialogue.}
  \label{eg-dialogue}
  \centering
\begin{tabularx}{\textwidth}{p{2cm}lllX}
\toprule
Role                  & Module & Slide Page & Timestamp & Message                                                                                                                                                                                           \\
\midrule 
AI Teacher            & 2      & 25         & 15:31     & This slide demonstrates the   parameter evolution and predictive performance of …                                                                                                                 \\
User (Human Student)  & 2      & 25         & 15:34     & I'm still confused about   this training process—could you explain it in more detail?                                                                                                             \\
AI Teacher            & 2      & 25         & 15:35     & Understanding the training   process of neural networks can be a bit complex, but don’t worry—we’ll take   it step by step. When gradient descent encounters multiple local minima...             \\
User (Human Student)  & 2      & 25         & 15:37     & Slide 23, in the steps of   stochastic gradient descent. Why update the parameters in the direction   opposite to the gradient                                                                    \\
AI Teacher            & 2      & 25         & 15:37     & When updating parameters, we   move in the opposite direction of the gradient because …                                                                                                           \\
User (Human Student)  & 2      & 25         & 15:39     & $\eta$ is the learning rate, which   controls the step size—right?                                                                                                                                    \\
Questioner            & 2      & 25         & 15:39     & Aha, well said! Tuning the   learning rate is definitely a bit of an art! The learning rate determines how   big a “step” we take during gradient descent. If the learning rate is too   small... \\
User (Human Student)  & 2      & 25         & 15:43     & Thinker, what do you think   of this issue?                                                                                                                                                       \\
Thinker               & 2      & 25         & 15:43     & Hmm, I think the selection of   learning rate is indeed quite subtle. If we start with a too high learning   rate …                                                                               \\
Questioner            & 2      & 25         & 15:43     & Wow, thinker, you brought up   a great question! The adaptive learning rate method is indeed cool as it can   …                                                                                   \\
Note Taker            & 2      & 25         & 15:43     & Thank you all for the   discussion! Let me briefly summarize the key points we talked about: …                                                                                                    \\
AI Teaching Assistant & 2      & 25         & 15:44     & The choice of learning rate   is indeed important, as it affects the convergence speed and performance of   the model…                                                                            \\
\bottomrule
\end{tabularx}
\end{table}

This study was conducted in the context of an introductory course entitled \textit{Towards Artificial General Intelligence}, offered on the MAIC platform. The course spanned eight weeks and comprised six distinct modules, each addressing a key aspect of artificial intelligence development. The six modules included: 1) Overview of General Artificial Intelligence; 2) Fundamentals of Neural Networks and Large Models; 3) Large Models Integrating Visual, Language, and Sound Inputs; 4) Autonomous Agents; 5) AI + X; and 6) AI Safety and Ethics. By systematically progressing through these modules, the course provided students with a comprehensive understanding of both the technical foundations and broader implications of general artificial intelligence. Since this course was at an introductory level, students learned relevant knowledge and completed after-course module quizzes, each containing several multiple-choice questions.

\subsection{Participants}

A total of 312 students (31.09\% female; age $Mean = 19.75$, $SD = 1.23$) from an elite university in China enrolled in the course on the MAIC platform. Participants were from various disciplines and voluntarily participated in the course. Ethical approval was granted by [UNIVERSITY NAME, anonymized for peer review). Consent forms were signed by all participants. To protect privacy, participants were assigned a unique user ID to access the system and complete questionnaires, thereby de-identifying their personal information. In the end, 110 students completed the entire course (30.91\% female; age $Mean = 19.96$, $SD = 1.18$). Both course completers and non-completers were included in the analysis.

\subsection{Procedure}

\begin{figure}[H]
  \centering
  \includegraphics[width=\textwidth]{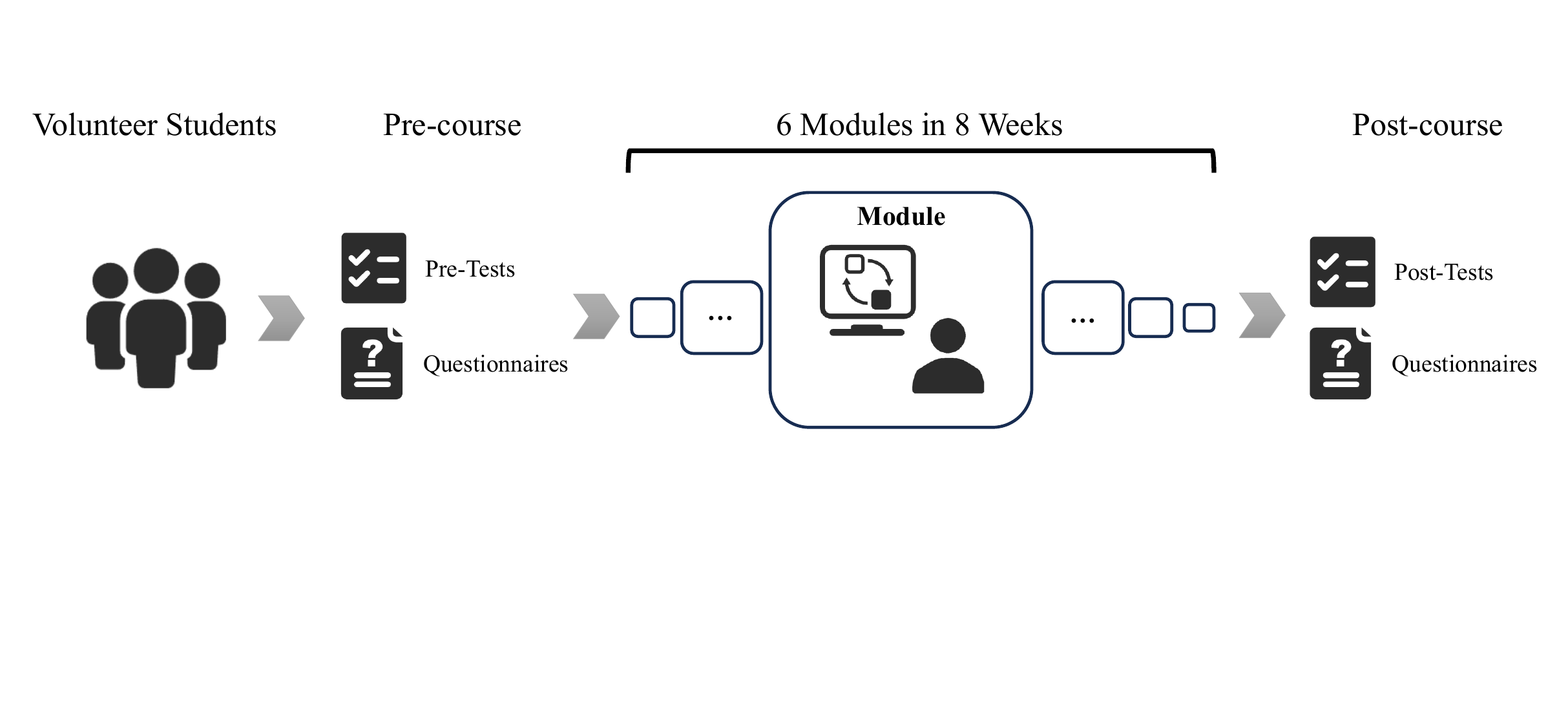}
  \caption{The experiment procedure.}
  \label{procedure}
\end{figure}

The study procedure is illustrated in Figure~\ref{procedure}. Prior to the course, all students were invited to complete pre-course questionnaires to gather demographic information (age, gender, and major), personality traits (Big-Five), non-cognitive skills (including academic self-efficacy and self-regulated learning), and their attitudes toward AI. Additionally, a pre-test was conducted to assess students’ prior knowledge relevant to the course content. The pre-test consisted of ten multiple-choice questions based on the learning materials, with a time limit of 15 minutes. Students were required to complete six modules in two months. Upon the completion of each module, they took a quiz consisting of ten multiple-choice questions. The learning process was self-paced, allowing students the flexibility to log in to the platform and complete the modules at their convenience. After completing the course, all students were asked to take a post-test, which included 60 multiple-choice questions. Students were also asked to fill out a series of questionnaires, which were identical to those used prior the study.

\subsection{Data collection}

The data for this study was obtained from multiple sources, including questionnaires (administered both pre- and post-course), system profiles, interaction log files, and multiple-choice tests (conducted before and after the course). All self- reported measures were designed using well- established scales with demonstrated reliability, as evidenced by high Cronbach’s $\alpha$ values (most values exceeding .8). The instruments used in this study include:

\paragraph{Big Five Personality Inventory}. The Big Five Personality Inventory is a prominent and widely accepted framework for understanding and measuring human personality traits, providing valuable insights into individual differences \citep{goldberg1993structure, john2008paradigm}. This model organizes personality into five essential dimensions: Neuroticism (N), Conscientiousness (C), Agreeableness (A), Openness to Experience (O), and Extraversion (E). The scale discussed here is derived from a shortened and culturally localized version, which includes three items for each of the five personality domains, making it efficient and adaptable for diverse populations \citep{zhang2019development}. Cronbach’s $\alpha$ coefficients are .87, .66, .87, .86, and .83 for N, C, A, O, and E, respectively.

\paragraph{Academic Self-Efficacy}. Academic self-efficacy refers to an individual’s belief in their ability to successfully complete a specific academic task or achieve a particular academic goal \citep{bandura1997self, eccles2002motivational, elias2002utilizing, linnenbrink2003role, schunk2002development}. In this study, academic self-efficacy was measured using an 8-item scale designed to assess students’ confidence in their academic performance \citep{chemers2001academic}. Cronbach’s $\alpha$ for this scale is .91.

\paragraph{Academic Learning Motivation}. Students’ academic learning motivation was evaluated using a 12 item scales including both self-determined motivation and non-self-determined motivation \citep{yu2018self}. Cronbach’s $\alpha$ for this scale is .80.

\paragraph{Attitudes toward AI}. Students’ attitudes toward AI stem from the broader concept of technology acceptance, which can be defined as "an individual’s psychological state regarding their voluntary and intentional use of a particular technology" \citep{masrom2007technology}. This study employs the UTAUT2 (Unified Theory of Acceptance and Use of Technology) model to examine such attitudes \citep{chang2012utaut, strzelecki2024students, venkatesh2012consumer}. A total of 30 items in the scale encompass eight core dimensions: Performance Expectancy, Effort Expectancy, Social Influence, Facilitating Conditions, Hedonic Motivation, Habit, Behavioral Intention, and Personal Innovativeness. Cronbach’s $\alpha$ for this whole scale is .95, with high validity across all internal dimensions (.91 for Performance Expectancy, .93 for Effort Expectancy, .86 for Social Influence, .85 for Facilitating Conditions, .92 for Hedonic Motivation, .87 for Habit, .92 for Behavioral Intention, and .89 for Personal Innovativeness respectively. 

\paragraph{Interaction logs}. Interaction logs were collected throughout the course, documenting conversation-based interactions between students and various agents. In total, 13,855 rounds of dialogue were recorded, comprising approximately 1.6 million Chinese characters. Each interaction entry included the speaker’s role, a timestamp, the utterance, and the corresponding slide page and module where the interaction took place.

\paragraph{Quizzes, pre- and post-tests}. Module quizzes and pre- and post-tests were created by the course preparation team (human course teacher and human teaching assistants) to assess students’ understanding of the course material. Each quiz included ten multiple-choice questions. The pre-test consisted of a total of ten items, while the post-test was more extensive and comprehensive, containing 60 items.

\subsection{Data analysis}
\label{sec:data-analysis}

\paragraph{Auto-encoding on student-AI conventional discourse}. To analyze the nature of students’ in-class interactions, an auto-encoding framework was developed to systematically categorize students’ messages. The coding scheme was designed to capture the nature of student-AI collaborative learning on three key dimensions: behavior, cognition, and emotion \citep{amatari2015instructional, flanders1967interaction}. The behavioral codes were primarily derived from the ten categories identified in Flanders’ Interaction Analysis System (FIAS), providing a structured foundation for classifying interaction types. The cognitive dimension was informed by Bloom’s Taxonomy \citep{anderson2001taxonomy, bloom1956handbook}, with cognitive levels categorized into three tiers: Remember \& Understand, Apply, and higher order thinking levels—Analyze, Evaluate \& Create. For the emotional dimension, codes were classified as positive, negative, or neutral, based on \citet{russell1980circumplex} circumplex model of affect. Each interaction assigned more than one codes to capture the nature of the discourse. To validate the framework, three senior education researchers conducted three rounds of trial coding on 200 dialogue entries, achieving an inter-coder reliability of 85\%. This indicates a high degree of agreement and consistency in the application of the coding scheme. The finalized code definitions are summarized in Table~\ref{autocoding}. The examples of coding are provided in Table~\ref{eg-coding}.

\begin{table}[H]
  \caption{Final code definitions.}
  \label{autocoding}
  \centering
  \begin{tabular}{lll}
    \toprule
    Dimension     & Code     & Description \\
    \midrule
    \multirow{8}{*}{Behaviour} & SB1  & Ask   questions                 \\
                           & SB2  & Respond   to questions          \\
                           & SB3  & Initiate   ideas                \\
                           & SB4  & Negotiate   and Confirm ideas   \\
                           & SB5  & Manage   classroom partners     \\
                           & SB6  & Monitor   and regulate progress \\
                           & SB7  & Share   emotion                 \\
                           & SB8  & Others                          \\
\midrule
\multirow{4}{*}{Cognition} & SC0  & Unrelated                       \\
                           & SC1  & Remember   \& Understand        \\
                           & SC2  & Apply                           \\
                           & SC3  & Analyze,   Evaluate \& Create   \\
\midrule
\multirow{3}{*}{Emotion}   & SE0  & Neutral                         \\
                           & SE1  & Positive                        \\
                           & SE2  & Negative      \\                 
    \bottomrule
  \end{tabular}
\end{table}

\begin{table}[H]
\caption{Example coding of student-AI interaction discourse.}
  \label{eg-coding}
  \centering
\begin{tabularx}{\textwidth}{p{2cm}Xp{1.5cm}p{1.5cm}p{1.5cm}}
\toprule
Role                  & Message                                                                                                                                                                                           & Behavior codes & Cognition codes & Emotional codes \\
\midrule
User (Human Student)  & I'm still confused about   this training process—could you explain it in more detail?                                                                                                             & SB5, SB1        & SC1             & SE1             \\
AI Teacher            & Understanding the training   process of neural networks can be a bit complex, but don’t worry—we’ll take   it step by step. When gradient descent encounters multiple local minima...             & /              & /               & /               \\
User (Human Student)  & $\eta$ is the learning rate, which   controls the step size—right?                                                                                                                                    & SB4            & SC1             & SE1             \\
AI teaching assistant & Aha, well said! Tuning the   learning rate is definitely a bit of an art! The learning rate determines how   big a “step” we take during gradient descent. If the learning rate is too   small... & /              & /               & /               \\
User (Human Student)  & What do you think about this   issue, Thinker?                                                                                                                                                    & SB6            & /               & SE1            \\
\bottomrule
\end{tabularx}
\end{table}

\begin{figure}[H]
  \centering
  \includegraphics[width=0.6\textwidth]{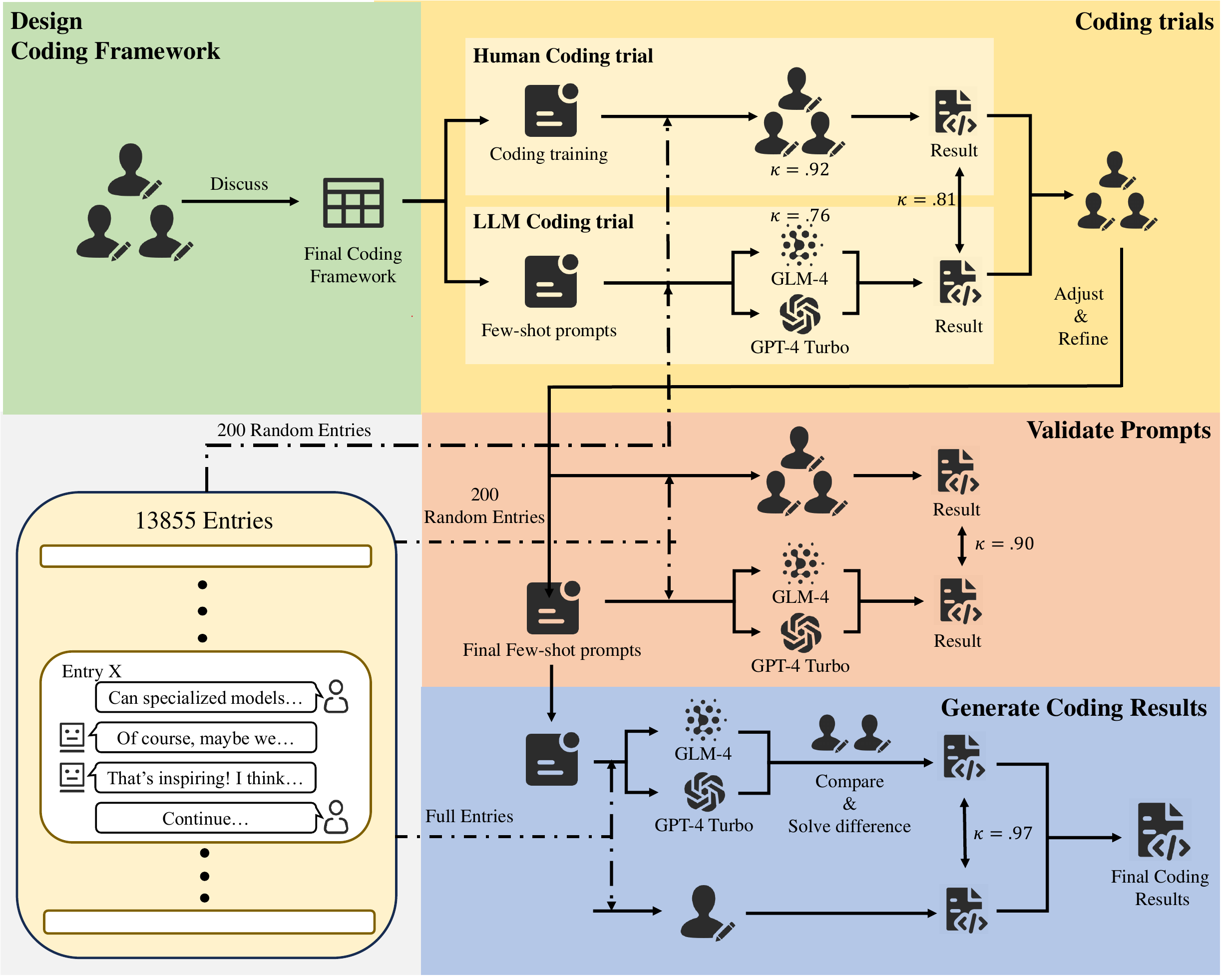}
  \caption{Coding process description.}
  \label{coding}
\end{figure}

Building on prior research demonstrating the utility of LLMs in supporting coding within educational contexts \citep{long2024evaluating}, this study explored the application of LLMs to assist in the coding process. Specifically, it leveraged two high-performing models, GPT-4o and GLM-4, and employed few-shot prompting techniques to achieve high accuracy. Human coders engaged in the loop to ensure the reliability of coding results. Figure~\ref{coding} describes the LLM-assisted coding process.

The coding process comprised three stages. In the first stage, a subset of 200 dialogue entries was manually coded and used to iteratively refine the prompts. Human coders continuously adjusted the prompts based on LLM- generated outputs, increasing the inter-rater reliability from 0.81 to 0.90 after multiple refinement cycles. In the second stage, to assess the generalizability of the refined prompts, another randomly selected set of 200 dialogue entries was coded independently by human coders and the two LLMs. The results demonstrated a reliability exceeding 0.90 across all dimensions and surpassing 0.95 in the cognitive and emotional dimensions, highlighting the prompts’ robustness and precision. In the final stage, the refined prompts were used to code the entire dataset. Two human coders then reviewed the outputs from both LLMs, while a third human coder independently coded the 30\% of data of the whole dataset. The final comparison yielded a reliability score of 0.97. Discrepancies were resolved through coder review and revision, producing the finalized coding results.

\paragraph{Data preparation}. To extract interaction data from the original log data, several aggregated variables were calculated using a Python script. These variables were measured at the student level and included metrics such as the total number of messages (MsgNum), the average message length (AvgMsgLen), the average quiz scores per module (AvgQuiz), and the percentage of each code (Pct\_[code]). Once these variables were computed for all students, the data was integrated with their pre- and post- test scores to create a comprehensive dataset.

\paragraph{Clustering analysis}. Clustering analysis can identify distinct student groups with varying engagement patterns, but the chosen dimensions for the analysis are critical to the clusters’ formation and interpretation. The primary objective of this study is to unearth the various learning behaviors exhibited by students during an online course. Thus, we firstly conducted a cluster analysis based on variables extracted from system profiles, including the total number of messages (MsgNum), average message length per message (AvgMsgLen), the number of completed Modules (CplModNum) and the total time students spent on the platform, measured in seconds (ModTime). The second cluster analysis was conducted using interaction characteristics extracted from student-AI dialogue records. The interaction characteristics contain all percentage of coded message contents (Pct\_[code]). Because the sum of percentage for each dimension of codes equals 1, the last code in each dimension (Pct\_SB8, Pct\_SC3 and Pct\_SE2) was omitted in the cluster process in order to avoid collinearity. Thus, the dimension count for the first cluster analysis is 4, and 12 for the second analysis.

Data pre-processing before clustering constitutes a key in the clustering process. For the numerical variables MsgNum, AvgMsgLen and ModTime, log transformation was employed to adjust their distributions. Followed by which, the z-score standardization was implemented to normalize all the dimension variables toward a common scale for both two cluster process.

Considering the data at hand, which contained 305 samples across 4 features for the first cluster analysis, 73 samples across 12 features for the second analysis, hierarchical clustering was determined to be the most suitable, due to its effectiveness in handling relatively smaller datasets and full capabilities to unearth underlying patterns among participants \citep{punitha2014performance, rana2016application}. This analysis utilized the NbClust package \citep{charrad2014nbclust} in R Studio to perform the pre-processing and hierarchal cluster analysis, selecting Ward’s method \citep{ward1963hierarchical}, a popular proximity matrix computation method as the dissimilarity measure \citep{le2023review}.

\paragraph{Epistemic Network Analysis}. Epistemic Network Analysis (ENA) is a specialized technique that assists in the recognition, quantification, and depiction of connections within coded data via undirected weighted network models \citep{shaffer2017epistemic}. This procedure capitalizes on the ability to visualize the intersecting nodes of codes generated from qualitative data. Furthermore, it enables the plotting of those connections within a two-dimensional space, courtesy of normalization and dimensional reduction techniques. Crucially, ENA also enables seamless comparative analysis between different groups \citep{shaffer2017quantitative}. This study focused on exploring differences among clusters and employs ENA on students’ message logs to discern their behavioral patterns. The data for this analysis is sourced from codified student message logs. The rENA package \citep{marquart2019rena} leveraged through R studio was used. 

\section{Results}

\subsection{Results of cluster analysis}
\label{sec:cluster1}

\begin{figure}[H]
  \centering
  \includegraphics[width=0.7\textwidth]{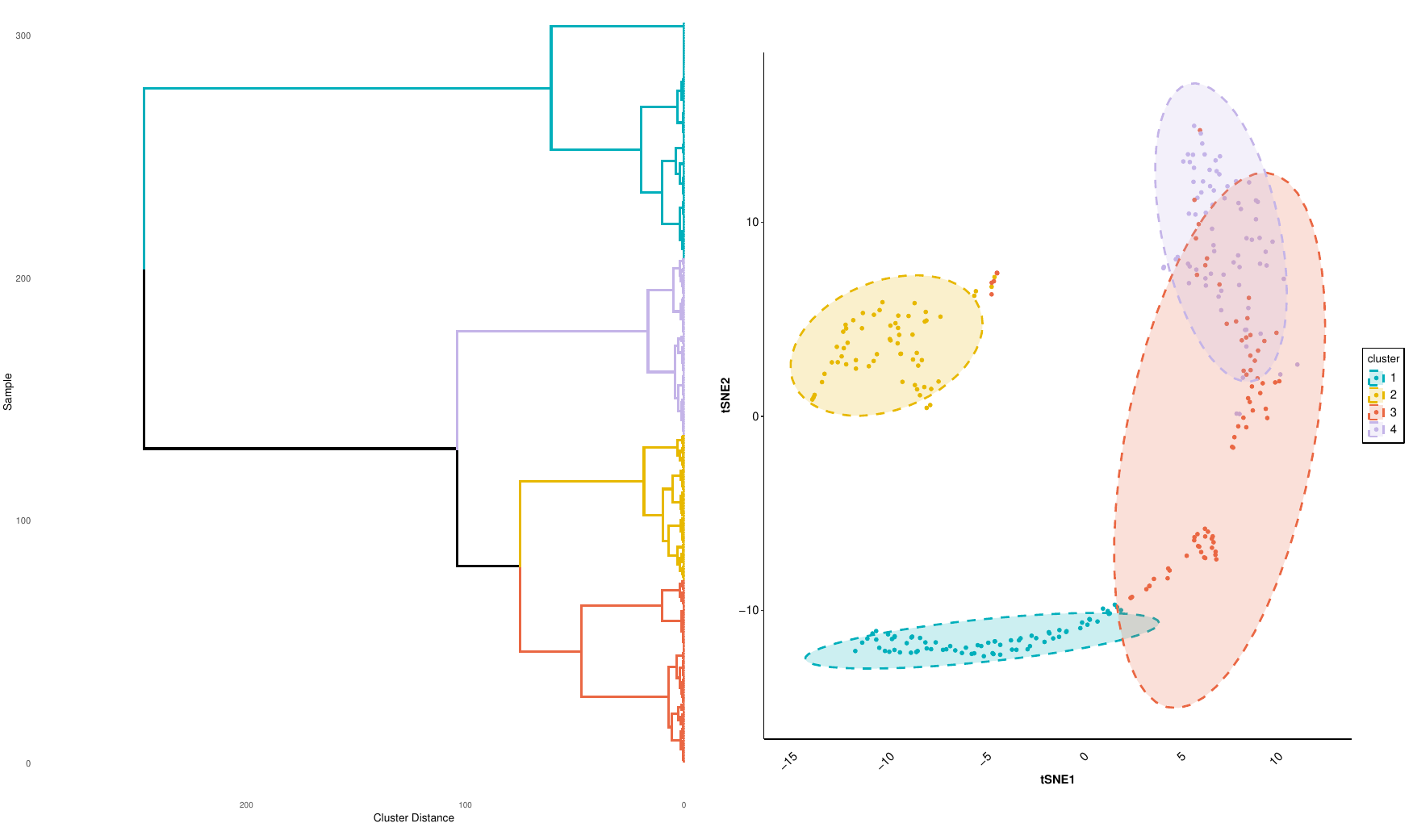}
  \caption{Dendrogram (left) and the t-SNE plot (right) of the course-level engagement cluster results.}
  \label{cluster1_den}
\end{figure}

We first clustered students' engagement in this multi-agent-driven course based on their course completion status, time spent on the course, number of messages, and the average length of each message. The hierarchical cluster method was applied using the Nbclust package in R, testing 23 distinct metrics, with the number of clusters ranging from 1 to 10. The majority decision, receiving 5 votes, suggested a cluster number of 4 for data segmentation. The four clusters consisted of 97, 60, 75 and 73 individuals, respectively. Dendrogram of the hierarchical cluster and the t-distributed Stochastic Neighbor Embedding (t-SNE) \citep{van2008visualizing} were shown in Figure~\ref{cluster1_den}.

\begin{figure}[H]
  \centering
  \includegraphics[width=0.7\textwidth]{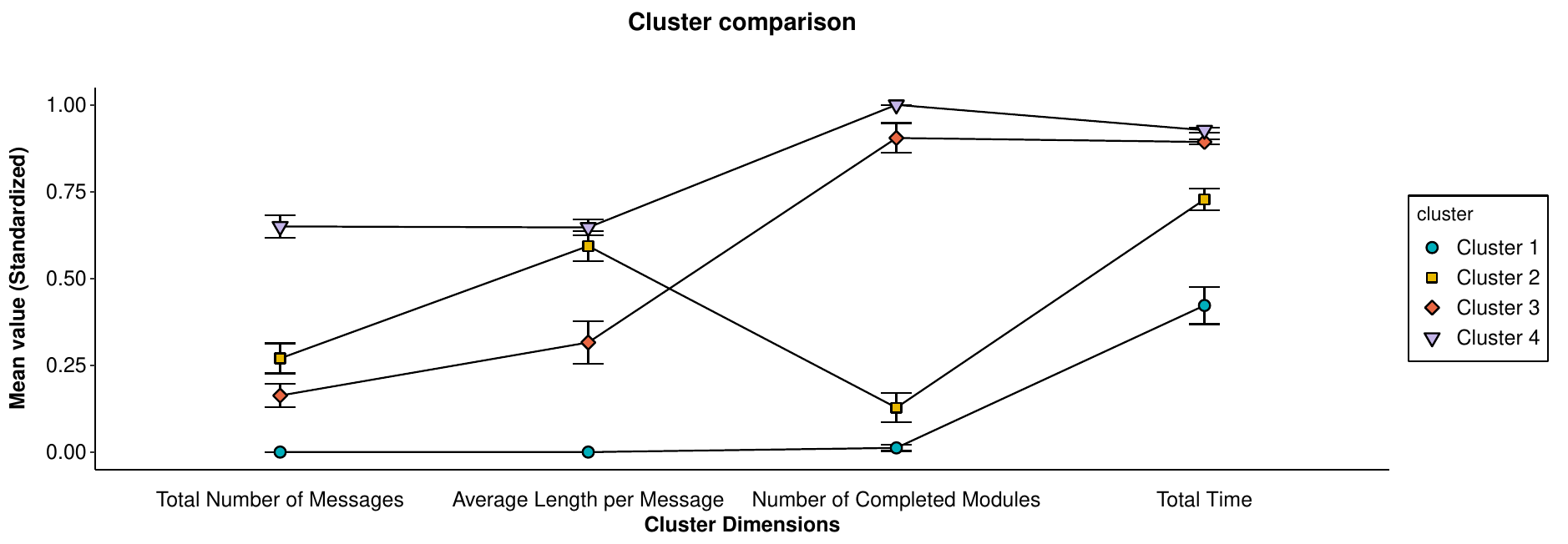}
  \caption{The distribution of four clusters across course engagement dimensions.}
  \label{cluster1_comp}
\end{figure}

To further explore the engagement differences among groups, we conducted non-parametric Kruskal-Wallis tests. This approach was chosen because Levene’s tests revealed significant unequal variances across the four groups ($p$s$ < .001$). The results suggested there were significant group differences on the total prompt number that students sent ($\chi^2 = 243.80$, $p <.001$), average message length ($\chi^2 = 213.18$, $p <.001$), completed module number ($\chi^2 = 233.50$, p <.001), and total time spent ($\chi^2 = 240.61$, $p <.001$). Distribution of four clusters across these dimensions was displayed in Figure~\ref{cluster1_comp}.

The four clusters displayed distinct engagement profiles:

Students in cluster 1 (Early Disengagers, n = 97) have no conversations with AI agents and did not complete any modules of this course. They spent the least amount of time on the system ($M = 1274.27$, $SD = 2508.20$). 

Cluster 2 (Stagnating Interactors, n = 60) students also showed a low course completion rate ($M = 0.77$, $SD = 1.00$) and spent significantly less time on the system ($M = 6495.45$, $SD = 6579.31$) compared to Clusters 3  ($M = 22891.41$, $SD = 7824.96$) and 4 ($M = 33306.27$, $SD = 11843.99$), $z = -5.77$, $p < .001$ and $z = -8.59$, $p < .001$, respectively. However, they sent a significantly higher number of messages ($M = 6.53$, $SD = 9.56$) than Cluster 1, $z = 8.50$, $p < .001$, and had a significantly greater average message length ($M = 23.79$, $SD = 19.25$) than Clusters 1 and 3 ($z = 11.0$, $p < .001$  and $z = 5.34$, $p < .001$, respectively). 

Cluster 3 students (Survived Lurkers, n = 75) had a high course completion rate ($M = 5.43$, $SD = 1.14$), significantly higher than students in Clusters 1 and 2 ($z = 12.5$, $p < .001$, and $z = 8.59$, $p < .001$, respectively). Notably, they sent a small number of messages ($M = 2.43$, $SD = 2.95$) with a relatively short average message length ($M = 8.86$, $SD = 12.70$). 

Finally, Cluster 4 students (Active Completers, n = 73) demonstrated the highest engagement. They had the highest course completion rate ($M = 6.00$, $SD = 0.00$), sent the most messages ($M = 46.81$, $SD = 42.15$), had the longest average message length ($M = 25.92$, $SD = 16.93$), and spent the most total time on the system ($M = 33306.27$, $SD = 11843.99$).

To investigate whether the differences in student course-level engagement among the four clusters were linked to individual characteristics, we analyzed pre-course questionnaire data on personality traits, learning motivation, and academic self-efficacy using ANOVA models. There were no significant differences found among the four clusters across the five Big Five personality dimensions: Neuroticism, $F(3, 301) = 0.14$, $p = .93$; Conscientiousness, $F(3, 301) = 0.28$, $p = .84$; Agreeableness, $F(3, 301) = 1.03$, $p = .38$; Openness to Experience, $F(3, 301) = 1.21$, $p = .31$; and Extraversion, $F(3, 301) = 1.14$, $p = .33$. Similarly, no significant differences were found among the four clusters for either learning motivation ($F(3, 301) = 0.78$, $p =.51$) or academic self-efficacy ($F(3, 301) = 0.74$ ,$p = .53$).

\begin{figure}[H]
  \centering
  \includegraphics[width=\textwidth]{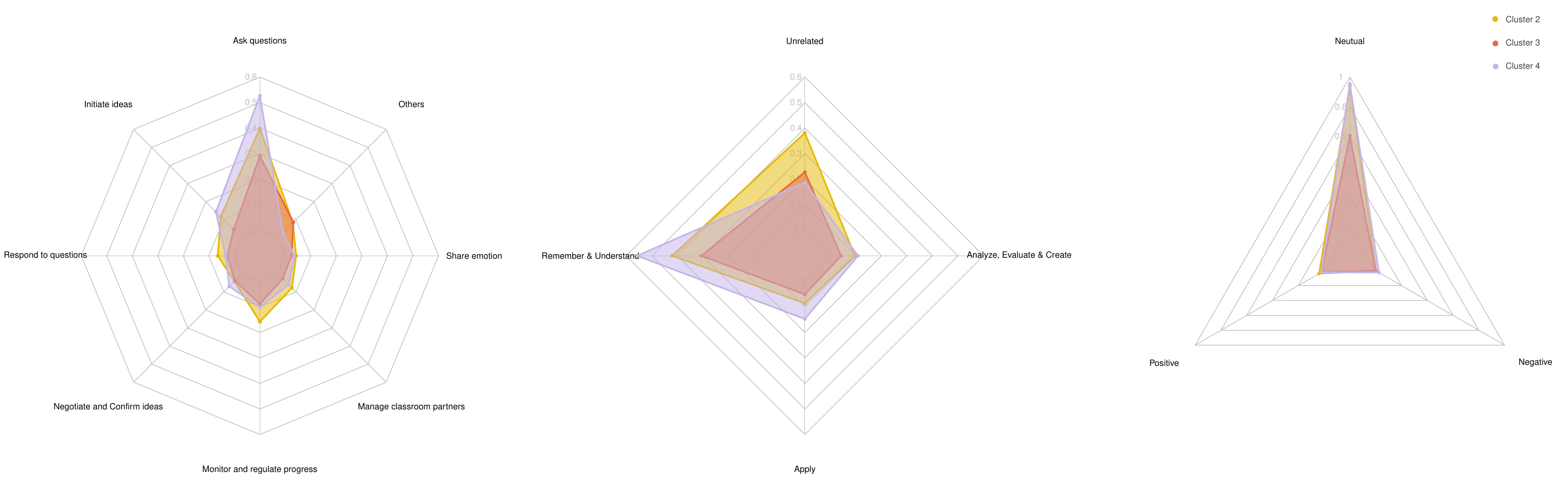}
  \caption{Radar plots among Cluster 2, 3, and 4 on message codes.}
  \label{cluster1_radar}
\end{figure}

Message codes were also examined across these clusters, except for Cluster 1 who did not send any messages during the course. Figure~\ref{cluster1_radar} displayed the radar plots of Cluster 2, 3 and 4 on the percentages of message codes. ANOVA tests were conducted to compare the distributions among the clusters, and significant differences were found in SC0 (Content Unrelated), $F(2, 205) = 6.10$, $p = .003$. Pairwise comparisons with Bonferroni adjustments demonstrated that Cluster 2 had higher unrelated contents ($M = 0.38$, $SD = 0.41$) than Cluster 3 ($M = 0.23$, $SD = 0.35$, $t = 2.67$, $p = .02$) and Cluster 4 ($M = 0.19$, $SD = 0.21$, $t = 3.35$, $p = .003$). Furthermore, there are also significant differences on SB1 (Ask questions), $F(2, 205) = 9.87$, $p < .001$. Cluster 4 had higher proportion of asking question ($M = 0.53$, $SD = 0.21$) than Cluster 2 $(M = 0.40$, $SD = 0.36$,$ t = 2.27$, $p = .07$) and Cluster 3 $(M = 0.29$, $SD = 0.37$,$ t = 4.44$, $p < .001$).

\subsection{Students' Engagement in Human-AI Interaction}

Our previous analysis of course engagement identified a group of students (Cluster 4, n = 73) who actively engaged in interactions with the AI agents. However, relying solely on message number and length was insufficient to reveal the distinct patterns of interaction within this group. To address this, we used the coding framework described in Section~\ref{sec:data-analysis} to code students' interactive messages with the AI agents. Each student message was coded based on its behavioral, emotional, and cognitive dimensions. Given the considerable variability in the number of messages sent by each student, we mitigated potential bias by using the proportion of each code category relative to the total messages. These proportions then served as the basis for a subsequent clustering analysis of students' human-AI interaction engagement. 

\begin{figure}[H]
  \centering
  \includegraphics[width=0.7\textwidth]{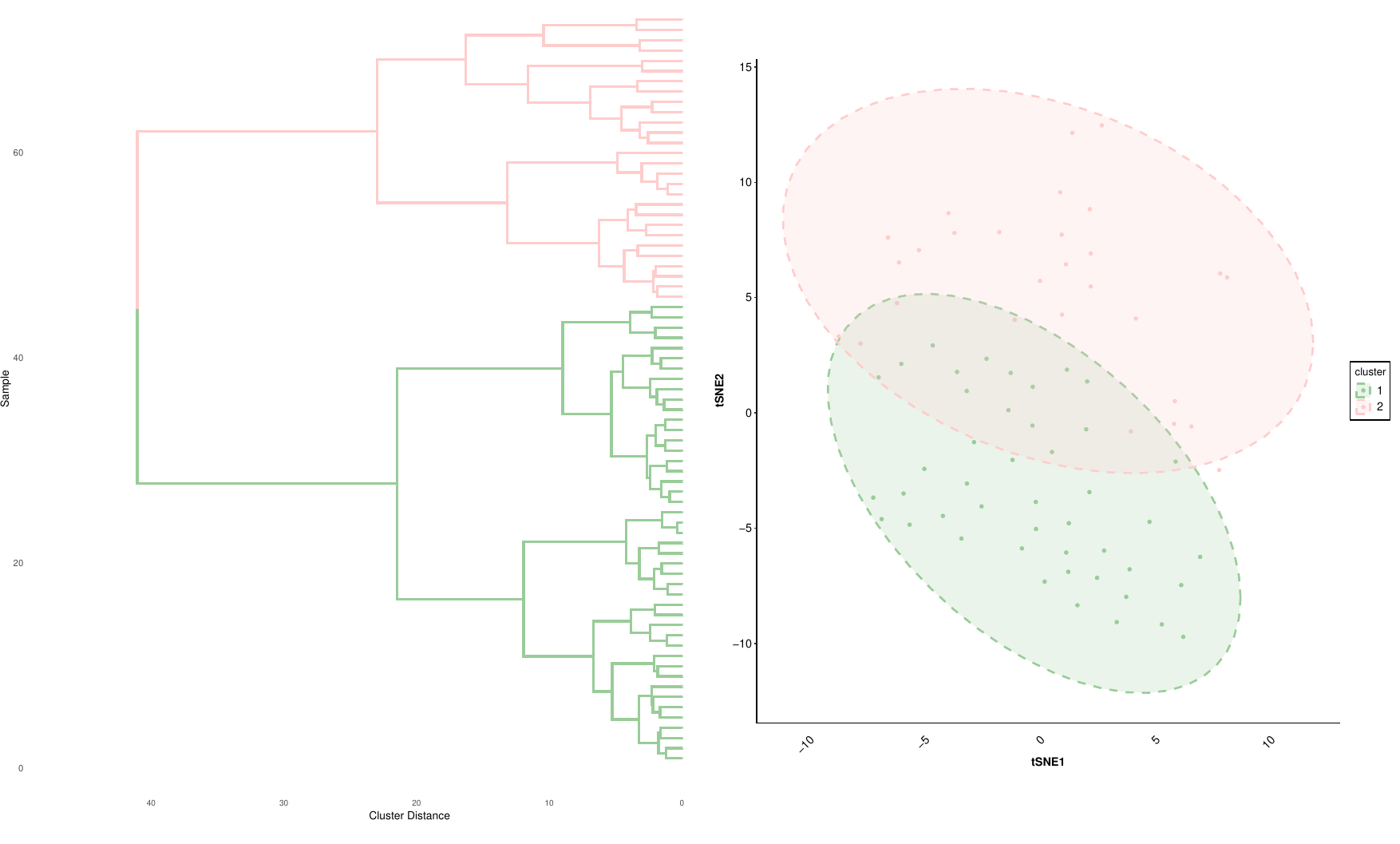}
  \caption{Dendrogram (left) and the t-SNE plot (right) of the cluster results based on dialogical codes .}
  \label{cluster2_den}
\end{figure}

Similar to Section~\ref{sec:cluster1}, we conducted a hierarchical cluster to test the 23 indices, and the results suggested a decision of 2 clusters with the most 5 votes. Therefore, seventy-three students in Cluster 4 were further categorized into two groups—Cluster 4–\Rmnum{1} and Cluster 4–\Rmnum{2}—with 45 and 28 students respectively. Dendrogram of the hierarchical cluster and the t-SNE visualization were shown in Figure~\ref{cluster2_den}.

To investigate the features of the two clusters, we conducted group-wise comparisons across the percentages of different codes on the three dimensions (behavioral, cognitive, and emotional). Preliminary tests (Shapiro-Wilk tests and Levene’s tests) were first conducted and failed for all variables, so non-parametric Mann-Whitney U tests were used. As shown in Table~\ref{tab:cluster2_comp}, the two subclusters had significantly different interaction patterns across the behavioral, cognitive, and emotional dimensions.

\begin{table}[]
\caption{Descriptive results and Mann-Whitney U tests results between 2 subclusters on percentages of message codes.}
\label{tab:cluster2_comp}
\centering
\begin{threeparttable}
\begin{tabular}{llcccccc}
\toprule
Dimension                  &     & \multicolumn{2}{l}{Cluster 4–\Rmnum{1}} & \multicolumn{2}{l}{Cluster 4–\Rmnum{2}} & \multicolumn{2}{l}{Mann-Whitney U tests} \\
                           &     & Mean           & SD             & Mean           & SD             & z              & p                       \\
                           \midrule
\multirow{8}{*}{Behavior}  & SB1 & 0.66           & 0.14           & 0.32           & 0.14           & \textbf{6.56 }          & \textbf{<.001}         \\
                           & SB2 & 0.01           & 0.02           & 0.07           & 0.09           & \textbf{-4.15}          & \textbf{<.001}         \\
                           & SB3 & 0.11           & 0.08           & 0.19           & 0.14           & \textbf{-2.36}          & \textbf{.02}                     \\
                           & SB4 & 0.08           & 0.07           & 0.05           & 0.04           & \textbf{2.39}           & \textbf{.02}                     \\
                           & SB5 & 0.07           & 0.09           & 0.15           & 0.18           & \textbf{-2.02}          & \textbf{.04}                     \\
                           & SB6 & 0.03           & 0.04           & 0.11           & 0.11           & \textbf{-3.31}          & \textbf{<.001}         \\
                           & SB7 & 0.02           & 0.07           & 0.07           & 0.06           & \textbf{-4.07}          & \textbf{<.001}         \\
                           & SB8 & 0.02           & 0.03           & 0.05           & 0.08           & \textbf{-2.28}          & \textbf{.02}                     \\
                           \midrule
\multirow{4}{*}{Cognitive} & SC0 & 0.11           & 0.12           & 0.30           & 0.27           & \textbf{-3.12}          & \textbf{.001}                    \\
                           & SC1 & 0.64           & 0.14           & 0.43           & 0.21           & \textbf{4.03}           & \textbf{<.001}         \\
                           & SC2 & 0.15           & 0.12           & 0.14           & 0.11           & 0.07           & .95                     \\
                           & SC3 & 0.10           & 0.09           & 0.12           & 0.14           & 0.03           & .97                     \\
                           \midrule
\multirow{3}{*}{Emotion}   & SE0 & 0.97           & 0.04           & 0.92           & 0.06           & \textbf{3.39}           & \textbf{<.001}         \\
                           & SE1 & 0.01           & 0.01           & 0.05           & 0.06           & \textbf{-3.90}          & \textbf{<.001}         \\
                           & SE2 & 0.02           & 0.04           & 0.03           & 0.04           & -0.98          & .33             \\
                           \bottomrule
\end{tabular}
\begin{tablenotes}
    \item Note: SB1: Ask questions; SB2: Respond to questions; SB3: Initiate ideas; SB4: Negotiate and confirm ideas; SB5: Manage classroom partners; SB6: Monitor and regulate progress; SB7: Share emotion; SB8: Others; SC0: Unrelated; SC1: Remember \& Understand; SC2: Apply; SC3: Analyze, Evaluate \& Create; SE0: Neutral; SE1: Positive; SE2: Negative
\end{tablenotes}
\end{threeparttable}
\end{table}

As shown in Table~\ref{tab:cluster2_comp}, in the dimension of behavioral engagement, the results showed that Cluster 4–\Rmnum{1} had a significantly higher proportion of asking question ($M = 0.66$, $SD = 0.14$) and negotiating and confirming ideas ($M = 0.08$, $SD = 0.07$) compared to Cluster 4–\Rmnum{2} ($M = 0.32$, $SD = 0.14$; $M = 0.05$, $SD = 0.04$, respectively), $z = 6.56$,$ p < .001$; and $z = 2.39$, $p = .02$, respectively. Cluster 4–\Rmnum{2} had a significantly higher proportion of responding to questions ($M = 0.07$, $SD = 0.09$) than Cluster 4–\Rmnum{1} ($M = 0.01$, $SD = 0.02$), $z = 4.15$, $p < .001$. In addition, Cluster 4–\Rmnum{2} also had higher proportions of regulatory behaviors including regulating process and managing classmates ($M = 0.15$, $SD = 0.18$, $and M = 0.11$, $SD = 0.11$, respectively) than Cluster 4–\Rmnum{1} ($M = 0.07$, $SD = 0.09$ and $M = 0.03$, $SD = 0.04$, respectively), $z = 2.02$, $p = .04$ and $z = 3.31$, $p < .001$ respectively. Cluster 4–\Rmnum{2} also had a greater proportion of emotion-sharing behaviors ($M = 0.07$,$ SD = 0.06$) compared to Cluster 4–\Rmnum{1} ($M = 0.02$, $SD = 0.07$), $z = 4.07$, $p < .001$.

Regarding the cognitive engagement, students in Cluster 4–\Rmnum{1} had higher proportions of cognitive activities at the level of Remember and Understand ($M = 0.64$, $SD = 0.14$) than those in Cluster 4–\Rmnum{2} ($M = 0.43$, $SD = 0.21$), $z = 4.03$, $p < .001$. In contrast, students in Cluster 4–\Rmnum{2} sent a significantly greater percentage of messages unrelated to the course content ($M = 0.30$, $SD = 0.27$) compared to students in Cluster 4–\Rmnum{1} ($M = 0.11$, $SD = 0.12$),$ z = 3.12$, $p = .001$. These content-unrelated messages included some off-topic chats, such as casual interactions with the AI, including questions like “Who are you?”, or “Teacher, do you have recommendations for lunch?”.

In addition, emotional engagement among the two groups were also compared. Both subclusters of Cluster 4 were similar in that their messages were predominantly neutral in emotion, with a mean proportion of 0.97 ($SD = 0.04$) for Cluster 4–\Rmnum{1} and 0.92 ($SD = 0.06$) for Cluster 4–\Rmnum{2}. Students in Cluster 4–\Rmnum{2} showed a significantly higher proportion of positive emotions ($M = 0.05$, $SD = 0.06$) than those in Cluster 4–\Rmnum{1} ($M = 0.01$, $SD = 0.01$), $z = 3.90$, $p < .001$. No significance was found in the proportion of negative emotions between the two subclusters ($z = -0.98$, $p = .33$).

\begin{figure}[]
  \centering
  \includegraphics[width=\textwidth]{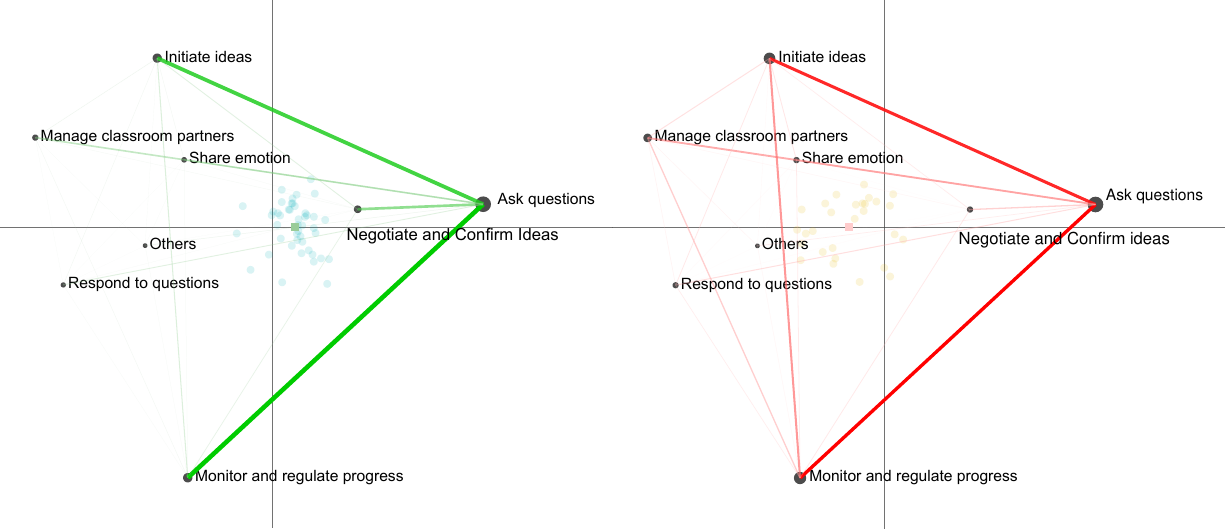}
  \caption{Cluster 4-\Rmnum{1} students’ (left) and Cluster 4-\Rmnum{2} students’ (right) networks.}
  \label{ena_mean}
  \caption*{Note: Centroids for each category are represented by solid squares, while individual student samples are depicted as semi-transparent circles.}
\end{figure}

To further investigate differences between two clusters in Cluster 4, inspecting the behavior modes of the 2 subgroups, ENA was conducted on the coding of students’ messages. The ENA model achieved a good fit of the data (Pearson correlation above 0.95). Along the x-axis (MR1), the t-test shows that Cluster 4–\Rmnum{1} ($M = 0.14$, $SD = 0.13$, n = 45) is statistically significantly different from Cluster 4–\Rmnum{2} ($M = -0.22$, $SD = 0.25$, n = 28; $t(40.26) = 8.47$,$ p < 0.001$, Cohen’s $d = 2.26$). Along the y-axis (SVD2), the t-test shows that there was no significant difference between the Cluster 4-\Rmnum{1} ($M = 0.00$, $SD = 0.25$, n = 45) and the Cluster 4–\Rmnum{2} ($M = 0.00$, $SD = 0.28$, n = 28), $t(51.78) = 0$, $p = 1.00$. 

Figure~\ref{ena_mean} illustrates the distribution of the two clusters’ centroids across the four quadrants in the network space. The placements of all nodes across the two clusters’ networks remained unchanged. The first quadrant is characterized by behaviors including asking questions and negotiating or confirming ideas, which highlight students’ cognitive engagement through joint knowledge construction. The second quadrant includes initiating ideas, managing peer interactions, and sharing emotions, reflecting a mode of socio-emotional engagement necessary to sustain collaborative learning. In contrast, the third quadrant is primarily associated with responding to questions and regulating progress, representing a supportive engagement focused on maintaining the flow and process of interactive learning. Taken together, this distribution illustrates the interplay of cognitive, socio-emotional, and regulatory dimensions in students’ messages, underscoring the multifaceted nature of student-AI interaction.

\begin{figure}[]
  \centering
  \includegraphics[width=0.5\textwidth]{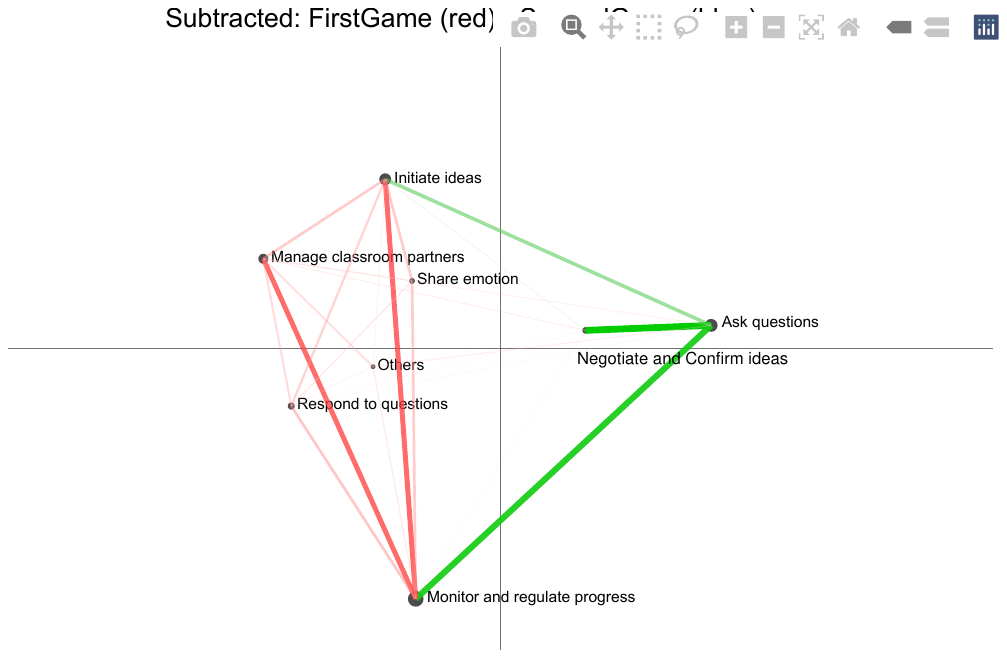}
  \caption{Subtraction Network between Cluster 4-\Rmnum{1} students (Green) and Cluster 4-\Rmnum{2} students (Red).}
  \label{ena_sub}
  \caption*{Note: Line colors indicate the relative co-occurrence of code pairs: greener lines represent higher co-occurrence in Cluster 4–\Rmnum{1}, while redder lines indicate higher co-occurrence in Cluster 4–\Rmnum{2}. }
\end{figure}

The elements in each quadrant collectively characterize the quadrant and the centroids located in the quadrant. The subtraction of the ENA networks of the two clusters (Figure~\ref{ena_sub}) further highlights the knowledge co-construction tendency of Cluster 4–\Rmnum{1} and the co-regulation tendency of Cluster 4–\Rmnum{2}. Students in Cluster 4–\Rmnum{1} made stronger connections to questioning and negotiating and confirming ideas. Their questioning behavior also had more frequent co-occurrence with initiate ideas, demonstrating that this group of students engaged in a collaborative, negotiation-based learning process. In comparison, students in Cluster 4–\Rmnum{2} made stronger interconnections among the other three core dimensions: progress regulation, peer management and emotion sharing. The denser network connections observed in both the second quadrant and the third quadrant suggest that this student cohort exhibits a pronounced propensity for learning regulation, indicating their tendency to monitor, control, and coordinate during their interactive learning with various AI agents. Based on their observed interaction styles with AI, the two clusters were labeled “active questioners” (Cluster 4–\Rmnum{1}) and "responsive navigators" (Cluster 4–\Rmnum{2}), respectively.

\subsection{How different engagement patterns associated with learning outcomes?}

This study further investigates the students’ learning outcomes, with particular attention to both knowledge acquisition and motivational changes. In terms of knowledge acquisition, descriptive results were listed in Table~\ref{tab:cluster1_gain}. An ANOVA test on pretest scores indicated that there were significant differences among four clusters ($F(3, 127) = 2.93$, $p = .04$). Further post-hoc tests showed that students in Cluster 4 ($M = 6.83$, $SD = 1.61$) obtained significantly higher pretest scores than Cluster 3 ($M = 5.96$, $SD = 1.46$), indicating that active course completers tend to have greater prior knowledge. When examining the subclusters, the ANOVA model showed that Cluster 4–\Rmnum{2} ($M = 7.19$, $SD = 1.68$) had significantly higher pretest scores than Cluster 3,$ t = 3.21$, $p = .005$. To examine students’ knowledge gain, average quiz scores and final exam scores were collected. Using pre-test results as the covariate, the ANCOVA test on average quiz scores showed a significant difference among Cluster 2, 3, and 4, $F(2, 125) = 7.20$,$ p = .001$. Cluster 1 was excluded from this comparison because these students did not complete any modules and thus had no quiz data. Post-hoc tests further revealed that students in Cluster 4 achieved significantly higher average quiz scores ($M = 7.18$, $SD = 1.23$) than those in Cluster 2 ($M = 6.00$, $SD = 1.63$) and 3 ($M = 6.07$, $SD = 1.40$), $t = 2.84$, $p = .02$ and $t = 3.23$, $p = .005$ respectively. Within the subclusters, Cluster 4-\Rmnum{1} obtained significantly higher average quiz scores ($M = 7.14$, $SD = 1.27$) than Cluster 2 and 3, $t = 2.82$, $p = .03$ and $t = 3.07$, $p = .01$ respectively. For final exam scores, the analysis was limited to students in Cluster 3 and Cluster 4, as they were the only ones who completed the exam. ANCOVA models did not show a significant difference in final exam scores between Cluster 3 and 4 ($F(1, 73) = 3.53$, $p = .06$) nor among Cluster 3 and two subclusters ($F(2, 72) = 2.62$, $p = .08$).

\begin{table}[H]
\caption{Descriptive results of pre- and post-test of knowledge across 4 clusters and 2 subclusters.}
\label{tab:cluster1_gain}
\centering
\begin{threeparttable}
    
\begin{tabular}{lllllll}
\toprule
                    & Cluster 1    & Cluster 2   & Cluster 3   & Cluster 4    &    &    \\
                    &              &             &             &              & Cluster 4–\Rmnum{1} & Cluster 4–\Rmnum{2}                                                  \\
                    \midrule
Pre-test            & \begin{tabular}[c]{@{}l@{}}6.50\\ (0.71)\end{tabular} & \begin{tabular}[c]{@{}l@{}}6.33\\ (2.01)\end{tabular} & \begin{tabular}[c]{@{}l@{}}5.96\\ (1.46)\end{tabular}  & \begin{tabular}[c]{@{}l@{}}6.83\\ (1.61)\end{tabular}  & \begin{tabular}[c]{@{}l@{}}6.62\\ (1.56)\end{tabular}  & \begin{tabular}[c]{@{}l@{}}7.19\\ (1.68)\end{tabular}  \\
Average quiz scores & /                                                           & \begin{tabular}[c]{@{}l@{}}6.00\\ (1.63)\end{tabular} & \begin{tabular}[c]{@{}l@{}}6.07\\ (1.40)\end{tabular}  & \begin{tabular}[c]{@{}l@{}}7.18\\ (1.23)\end{tabular}  & \begin{tabular}[c]{@{}l@{}}7.14\\ (1.27)\end{tabular}  & \begin{tabular}[c]{@{}l@{}}7.26\\ (1.17)\end{tabular}  \\
Final course exam   & /                                                           & /                                                           & \begin{tabular}[c]{@{}l@{}}35.18\\ (7.16)\end{tabular} & \begin{tabular}[c]{@{}l@{}}39.05\\(6.55)\end{tabular} & \begin{tabular}[c]{@{}l@{}}39.93\\(5.58)\end{tabular} & \begin{tabular}[c]{@{}l@{}}37.48\\ (7.91)\end{tabular} \\
\bottomrule
\end{tabular}
\begin{tablenotes}
    \item Note: Values are presented as means, with standard deviations in parentheses.
\end{tablenotes}
\end{threeparttable}
\end{table}

\begin{figure}[H]
  \centering
  \includegraphics[width=0.8\textwidth]{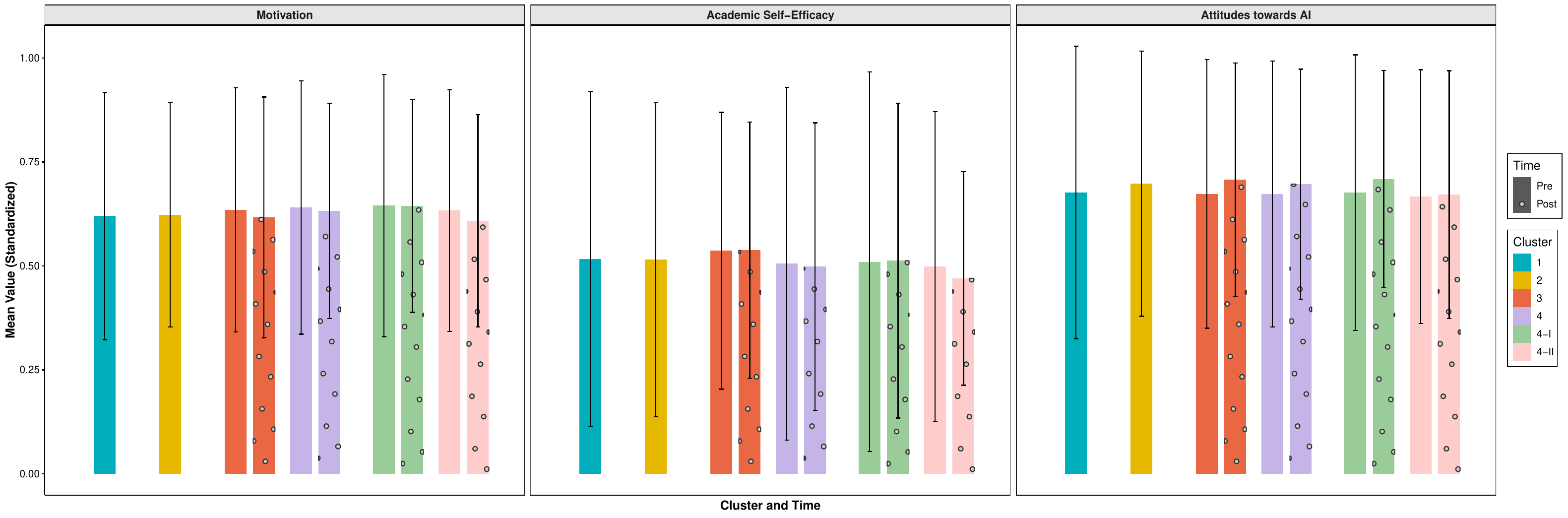}
  \caption{Results of pre- and post-tests of Motivation, Academic Self-Efficacy and Attitudes towards AI across 4 clusters and 2 subclusters.}
  \label{fig:cluster1_non_gain}
  \caption*{Note: Post-tests for Cluster 1 and 2 are missing data, represented as blank bars.}
\end{figure}

For non-cognitive traits, differences between pre- and post- questionnaire scores within each cluster were firstly examined using the paired t-test and the non-parametric Wilcoxon method when normality or homogeneity of variances was violated. The Wilcoxon test indicated a significant increase in attitudes towards AI in Cluster 4 ($z = 2.31$, $p = .02$) and a potentially significant increase in Cluster 3 ($z = 1.93$, $p = .05$). No significant changes were detected between the pre- and post- questionnaire on motivation and academic self-efficacy across all 4 clusters and 2 subclusters. Regarding the between-cluster comparisons on affective changes, using pre-questionnaire data as the covariate, the ANCOVA models indicated that there were no significant differences among clusters across Motivation ($F(2, 104) = 0.22$, $p = .80$), Academic Self-Efficacy ($F(2, 104) = 0.99$, $p = .37$) and Attitudes towards AI ($F(2, 104) = 0.63$, $p = .53$). Results were plotted in Figure~\ref{fig:cluster1_non_gain}.

\section{Discussion}

This study focuses on understanding the diverse learning engagement patterns of students in a human-AI collaborative learning environment. Through a two-step cluster analysis, based on both course-level engagement data and detailed student-AI interaction data, our research identified two distinct types of course non-completers and three types of course completers. These findings shed light on the complex nature of engagement and provide a nuanced view of student behavior in technology-enhanced learning environments.

\subsection{Diverse Engagement Patterns and Dropout Behaviors}

Our analysis identified two distinct types of students who dropped out of the course. The first group, comprising roughly 31\% of participants, exhibited a pattern of early disengagement. They signed up and briefly browsed the system before quickly dropping the course. In contrast, another 19\% of students initially invested in the course, actively interacting with the AI agents before disengaging and failing to complete all modules. While these dropout rates may seem significant, they fall within an acceptable range when compared to many traditional MOOC environments, where completion rates are often reported to be below 5\% \citep{onah2014dropout, kizilcec2013deconstructing, seaton2013does}. This suggests that the interactive and supportive nature provided by the multi-agent-empowered learning system may have helped retain a larger portion of students.

Similarly, among the completers of the course, our study identified three distinct groups with different interaction styles. The first group, termed "active questioners", actively engaged in interactions, with a high frequency of questioning and negotiation behaviors. The second, referred to as "responsive navigators", also actively participated but with a significantly different engagement style, often displaying responses to AI agents and regulatory behaviors. The third type of completer we identified were “lurkers”, who barely engaged in interactions with the AI agents but successfully completed the course. These findings provide a valuable supplement to existing research, aiding in a deeper understanding of how learners engage within highly interactive, AI-supported environments.

It is noteworthy that our study did not reveal any clear distinguishing, pre-existing traits between the dropout and completer groups. We explored several variables commonly associated with engagement in the traditional online learning literature, including prior knowledge, the Big Five personality traits, learning motivation, and self-efficacy, as well as students’ attitudes toward AI. However, we found no significant differences on these variables between students who dropped out and those who completed the course. This may be due to the homogeneity of our participant group, as most were low-level undergraduates from the same university in China with similar levels of motivation and self-efficacy. Within this group, dropout behavior may have been more random. Furthermore, this finding suggests that other unmeasured factors may influence engagement. For example, we did not investigate students' perceived course difficulty or the influence of co-learning relationships, both of which have been linked to engagement in previous studies \citep{breslow2013studying, qiu2016modeling}. Future research should investigate how these unmeasured factors might be associated with engagement in AI-instructed learning environments.

\subsection{Decoupling Interaction from Engagement in AI-Supported Learning}

In this multi-agent-empowered learning environment, our findings suggest that while student interaction and engagement are closely intertwined, they are not always entirely synonymous. Interaction often provides a behavioral trace of engagement, yet it fails to capture the construct’s full complexity. By examining distinct clusters of learners, we have observed clear evidence of both alignment and misalignment between these two dimensions. For example, students in Cluster 1 displayed no interaction and failed to complete any modules, representing a clear state of disengagement. Conversely, those in Cluster 4 exhibited frequent interaction with AI agents alongside high course completion, representing deep and effective engagement. These patterns align with prior learning analytics studies that equate behavioral activity with engagement \citep{wilson2021mapping, pazzaglia2016analysis, esnashari2018clustering}.

However, the contrasting profiles of Cluster 2 and Cluster 3 challenge this assumption and reveal the complexity of the relationship between interaction and engagement. Students in Cluster 2 showed a high volume of interaction with the AI agents but did not complete the course. This suggests that a high message count may not always signify productive engagement; instead, their frequent interactions might reflect a curiosity about the novel learning format itself, rather than a genuine interest in the course content. As a result, these students engaged in significantly more off-topic queries and exploratory behaviors, which were behaviorally active but ultimately unsuccessful in leading to course completion. This pattern highlights a form of engagement that is not directly tied to the learning content and proves to be unproductive in achieving learning goals. Conversely, Cluster 3 students demonstrated silent yet successful course completion, and their final exam performance was not different to that of active students in Cluster 4. This points to a different learning style, where students were capable of independently navigating the course material and only utilized the AI agents when absolutely necessary. This behavior aligns with findings from prior research that identified similar “lurker” patterns among learners \citep{milligan2013patterns}. Their low interaction levels did not hinder their learning, suggesting a high degree of cognitive engagement that was not outwardly visible through dialogue. These findings underscore that interaction frequency alone can be misleading, and that a deeper understanding of interaction quality, purpose, and learner strategy is essential for capturing the multifaceted nature of engagement. 

In addition, we found that students’ prior knowledge level may have been a key factor influencing their AI interaction patterns and learning outcomes. Students in Cluster 4, who had more interaction with the AI, obtained significantly higher pretest scores than those in Cluster 3, who barely interacted. This suggests that active interaction with AI may require a certain level of prior knowledge. Cognitive Load Theory provides a possible explanation: when students have a high level of prior knowledge, the intrinsic cognitive load required to process new information is significantly reduced. This allows them to quickly match and integrate new knowledge with existing schema rather than starting from scratch. This low-load state gives them more cognitive resources to perform higher-order learning activities, such as complex interaction, deep questioning, negotiation, and joint knowledge construction with the AI.

\subsection{Constructive Interaction Enhances Short-Term Learning Outcomes}

Our analysis further examined how different patterns of interaction among course completers relate to learning outcomes. Students in Cluster 4–\Rmnum{1}, who actively engaged in knowledge negotiation and co-construction with AI agents, achieved significantly higher average quiz scores than their counterparts in Cluster 3, who completed the course with minimal interaction. This suggests that constructive engagement, rather than passive reception, plays a crucial role in promoting deeper cognitive processing and knowledge internalization. Actively externalizing one’s thinking through dialogue with AI may enhance memory consolidation and conceptual understanding, thereby yielding improved performance on assessments targeting recent knowledge acquisition \citep{liu2020application, tang2025generative}. However, no significant differences were observed between Cluster 4-\Rmnum{1} and Cluster 4-\Rmnum{2} in quiz scores, despite the latter displaying higher levels of regulatory behaviors and positive emotions. This may be attributed to differences in prior knowledge, as students in Cluster 4-\Rmnum{2} may have found the learning content more accessible and thus focused more on emotional regulation or strategic navigation than on content negotiation. Importantly, this alternative interaction style did not appear to hinder learning outcomes.

Interestingly, final exam performance did not significantly differ across the three clusters. One possible explanation is that the quizzes assessed short-term mastery of discrete knowledge points more directly impacted by AI-mediated interaction, while the final exam may have required broader conceptual integration and long-term retention, which are less sensitive to short-term interaction patterns. Alternatively, the non-significant differences may stem from methodological limitations: since only a subset of students completed the final exam, the reduced sample size may have limited statistical power to detect meaningful group differences.

Beyond learning performance, our data revealed that all course completers—regardless of their interaction style—reported significantly more positive attitudes toward AI technology by the end of the course compared to baseline. This finding suggests that sustained engagement in a multi-agent-supported learning environment fosters increased acceptance and appreciation of AI in education. However, no significant changes were found in students’ learning motivation or academic self-efficacy, indicating that short-term participation in such environments may not be sufficient to shift more stable individual traits \citep{yan2025complexity, fan2025beware}.

\subsection{Implications}

This study offers several important implications for the design and evaluation of AI-supported learning environments. First, our findings highlight the need to move beyond simple behavioral indicators—such as clickstream, message frequency—to more nuanced measures of engagement that consider the quality, purpose, and outcomes of student–AI interaction. From a methodological standpoint, our clustering approach is a notable contribution as it leverages dialogue mining to understand engagement, an innovation beyond traditional online learning analytics that often relies on clickstream data or participation status in online discussion. Our findings also highlight that we should be cautious in equating high levels of interaction with effective learning, as excessive but unproductive engagement may signal wandering or struggle rather than success.

Second, the superior performance of students who engaged in constructive knowledge-building dialogues with AI agents underscores the value of designing AI systems that not only respond to queries but also promote deep cognitive engagement through prompting, elaboration, and negotiation. Embedding scaffolding strategies that encourage students to explain, reflect, or justify their thinking may help translate surface-level activity into meaningful learning gains.

Third, the observed positive shift in students’ attitudes toward AI suggests that sustained exposure to multi-agent systems can enhance learners’ openness to educational technologies. This may have long-term benefits in fostering digital readiness and reducing resistance to AI integration in education. However, the lack of change in students’ motivation and self-efficacy points to the limits of short-term interventions and highlights the importance of longitudinal designs and more personalized support mechanisms.

\subsection{Limitations}

Several limitations of this study should be acknowledged. First, the sample was drawn from a relatively homogenous group of undergraduate students from a single institution in China, which may limit the generalizability of the findings to more diverse educational contexts or learner populations with varied cultural, academic, or technological backgrounds. Second, while our clustering approach revealed meaningful patterns of interaction and engagement, the interpretation of these clusters relied on behavioral log data, which—despite its granularity—cannot fully capture students’ cognitive or emotional states during learning. Complementary data sources such as self-reports, interviews, or multimodal indicators (e.g., eye-tracking, EEG) could provide richer insight into the learner experience. Finally, this study focused on short-term engagement patterns within a single course context. As such, it remains unclear how interaction styles evolve over time, or how sustained exposure to AI agents might influence learning trajectories, academic beliefs, or self-regulated learning in the long run. Future research should address these limitations by expanding the participant pool, incorporating longitudinal and multimodal designs, and triangulating interaction data with richer contextual and psychological measures.

\section{Conclusion}

This study provided a multifaceted analysis of student engagement in a human-AI interactive learning environment. Our findings highlight the complex and diverse nature of learner engagement, moving beyond a simple dichotomy of active versus passive participation. By identifying distinct learner profiles—including Early Disengagers, Stagnating Interactors, and successful Lurkers—we demonstrated a crucial decoupling of behavioral interaction from meaningful learning engagement. Specifically, the results showed that high levels of interaction do not always equate to productive learning, as exemplified by students who frequently asked off-topic questions and ultimately dropped out. Conversely, a lack of overt interaction does not necessarily indicate disengagement, as evidenced by successful lurkers.

Our research also revealed that students’ prior knowledge significantly influenced their interaction patterns and short-term learning outcomes, suggesting that a foundational understanding may be a prerequisite for deep, constructive engagement with AI. However, we found no clear relationship between pre-existing learner traits and course completion, pointing to the influence of other unmeasured contextual factors. These findings have important implications for the design of adaptive educational systems. By understanding these nuanced engagement patterns, educators and designers can move beyond simple interaction metrics to create systems that provide timely, targeted support for struggling learners and recognize the silent yet successful engagement of others. Future research should investigate these unmeasured factors and explore longitudinal engagement patterns to gain a more complete understanding of learning in AI-assisted environments.

\begin{ack}
This work was funded by the Beijing Educational Science Foundation of the Fourteenth 5-year
Planning (BAEA24024).
\end{ack}

{
\small
\bibliography{ref}
}

\end{document}